\documentclass{ws-p8-50x6-00}

\begin{document}

\title{Parton Distributions in Nuclei at Small x}

\author{Ina Sarcevic}

\address{Department of Physics, University of Arizona, Tucson, AZ 85721, 
USA
\\E-mail: ina@gluon.physics.arizona.edu}


\maketitle

\abstracts{
We study the nuclear shadowing effect in the context of Glauber-Gribov 
multiple-scattering model and perturbative QCD.  We find that at small x, 
the $Q^2$ evolution of the shadowing is much slower than 
the DGLAP evolution, due to the multiple scatterings at small x.  
We show that the gluon shadowing at small x and for $Q^2 > 3GeV^2$ is 
perturbative in nature and does not depend on the initial, non-perturbative 
condition.  We evaluate the impact parameter dependence of the gluon distribution  
and show that it is a non-linear effect in the nuclear thickness function.}


The semihard processes, 
such as minijets, dileptons, open charm and direct photon 
production in heavy-ion collisions at $\sqrt{s}\geq 200$ GeV, can be 
reliably calculated in the framework of
perturbative QCD assuming 
the validity of the
factorization theorem in perturbation theory,
and with the knowledge of 
parton distributions in nuclei at small x.  

Quark distribution in nuclei has been measured and the attenuation 
has been firmly established
experimentally at CERN \cite{cern} and Fermilab \cite{fermilab} 
in the region of small $x$ 
in deeply-inelastic lepton scatterings (DIS) off nuclei.  
The data, taken over a
wide kinematic range, $10^{-5}<x<0.1$ and 0.05 ${\rm GeV}^2 <
Q^2< 100$ GeV$^2$, show a systematic reduction of nuclear structure function
$F_2^A(x,Q^2)/A$ with respect to the free nucleon structure function
$F_2^N(x,Q^2)$.  However, there is no experimental measurement of the 
gluon density in nuclei.  

At low $Q^2$, in DIS, the interaction of the virtual 
photon with the nucleons in the rest frame of the 
target is most naturally 
described by a vector-meson-dominance (VMD) model \cite{vmd}.  
At $Q^2>1\sim 2$ GeV$^2$, the virtual photon can penetrate the
nucleon and probe the partonic degrees of freedom where
 a partonic interpretation based on perturbative QCD is most relevant
in the infinite momentum frame.  
In the target rest frame, the virtual photon interacts with nucleons
via its quark-antiquark pair ($q\bar q$) color-singlet 
fluctuation \cite{lu}.  
If the coherence
length of the virtual photon is larger than the distance between 
nucleons in a nucleus, 
$l_c>R_{NN}$, the $q\bar q$ configuration interacts
coherently with fraction of the nucleons, while for 
$l_c>R_A$ (i.e. $x < 10^{-2}$), 
it interacts coherently with all the nucleons, with a 
cross section given by the
color transparency mechanism for a point-like color-singlet configuration, 
$\sigma_{q\bar q N} = \frac
{4\pi^2}{3} 
r_t^2\alpha_sx'g_{\rm DLA}(x',1/r_t^2)$ 
\cite{strikman}.  This results in a 
reduction of the total cross section and consequently attenuation of the 
parton distributions in nuclei.  

In the Glauber-Gribov multiple-scattering theory \cite{glauber}
nuclear collision
is a succession of collisions of the probe 
with individual nucleons within nucleus.
A partonic system ($h$), being a $q\bar q$ or a $gg$ fluctuation, 
can scatter coherently from several or all nucleons during its passage
through the target nucleus. The interference between the multiple scattering
amplitudes causes a reduction of the $hA$ cross section compared
to the naive scaling result of $A$ times the respective $hN$ cross 
section, the origin of the nuclear shadowing.  
The total $hA$ cross section is given by 
\begin{eqnarray}
\sigma_{hA}=\int d^2{\bf b} 2 
\left [ 1-e^{-\sigma_{hN}T_A({\bf b})/2}\right ]  
=  2\pi R_A^2\left [ \gamma +\ln \kappa_h +E_1(\kappa_h )\right ]\; ,
\end{eqnarray}  
where 
$\kappa_h =A\sigma_{hN}/(2\pi R_A^2)$ 
is an impact parameter averaged effective number of scatterings.  
For small value of $\kappa_h$, 
$\sigma_{hA}\rightarrow 2\pi R_A^2\kappa_h =A\sigma_{hN}$, the total $hA$
cross section is proportional to $A$.  In the 
limit $\kappa_h \rightarrow \infty$, the destructive interference between
multiple scattering amplitudes reduces the cross section, 
$\sigma_{hA}\rightarrow 2\pi R_A^2 (\gamma +\ln \kappa_h )$. Namely,
the effective number of scatterings is large and 
the total cross section approaches the geometric limit $2\pi R_A^2$, a
surface term which varies roughly as $A^{2/3}$. 

In the Glauber-Gribov eikonal approximation, 
\begin{eqnarray}
\sigma (\gamma^\star A)
=
\int d^2 b
\int_0^1dz \int d^2{\bf r}|\psi (z,{\bf r})|^2
2 (1-e^{-\sigma_{q\bar q N}({\bf r},z) T_A(b)/2}),
\end{eqnarray}
where
$|\psi (z,{\bf r})|$ 
is the photon wave function \cite{agl} and 
$z$ is the fraction of the energy carried 
by the quark (antiquark).  
The nuclear cross section is therefore reduced  
when compared to the simple addition of free nucleon cross sections.

At small $x$, the structure function of a nucleus, 
$F_2^A(x,Q^2)$ can be obtained from 
$\sigma (\gamma^*A)$.  
Substituting integration over ($z, {\bf r_t}$) to ($x',
Q'^2$) in (2), one obtains
for (sea)quarks \cite{us} 
\begin{eqnarray}
xf_A(x,Q^2) & = & xf_A(x,Q_0^2)+\frac{3R_A^2 x}{8\pi^2}\int_x^1 \frac{dx'}{x'^2}
\int_{Q_0^2}^{Q^2}
dQ'^2 
 \left [ \gamma +\ln (\kappa_q) +E_1(\kappa_q )
\right ]\; ,\label{eq:fa}
\end{eqnarray}
where 
$\kappa_q(x,Q^2) 
=\frac{2A\pi}{3R_A^2Q^2}\alpha_s(Q^2)xg_N^{\rm DLA}(x,Q^2)
$.  
We find 
our result for $F_2^N(x,Q^2)$ to be in excellent agreement with the 
recent HERA data in the kinematic region of $x$ and $Q^2$ 
where DLA is valid \cite{us}.  

We parametrize the 
initial shadowing ratio $R_0^q(x)\equiv F_2^A(x, Q_0^2)/
A F_2^N(x, Q_0^2)$ 
at $Q_0^2=0.4$ GeV$^2$ using 
experimental data and 
we calculate the nuclear structure function at the measured
$\langle Q^2\rangle$ values at different $x$ values. 
The results for $^{40}$Ca and $^{208}$Pb are shown in
Fig. 1.  
\begin{figure}[tbhp]
\vspace{-0.2cm}
$$
\begin{array}{c}
    \epsfxsize=6.4cm 
    \epsfbox{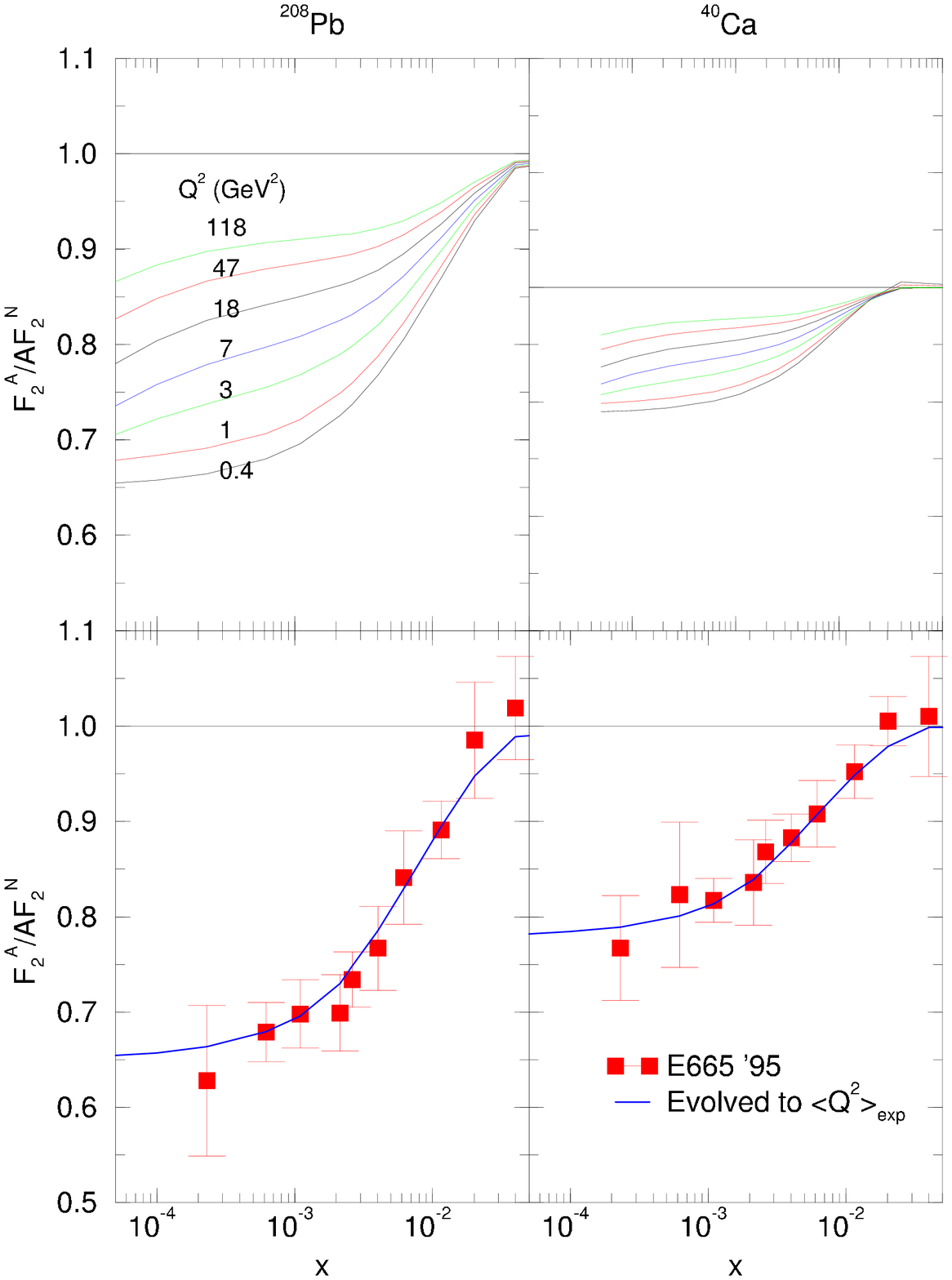}\\
    \rm{Fig.~1}
\nonumber
\end{array}
$$
\end{figure}
\begin{figure}[tbhp]
\vspace{-0.2cm}
$$
\begin{array}{c}
    \epsfxsize=8.4cm 
    \epsfbox{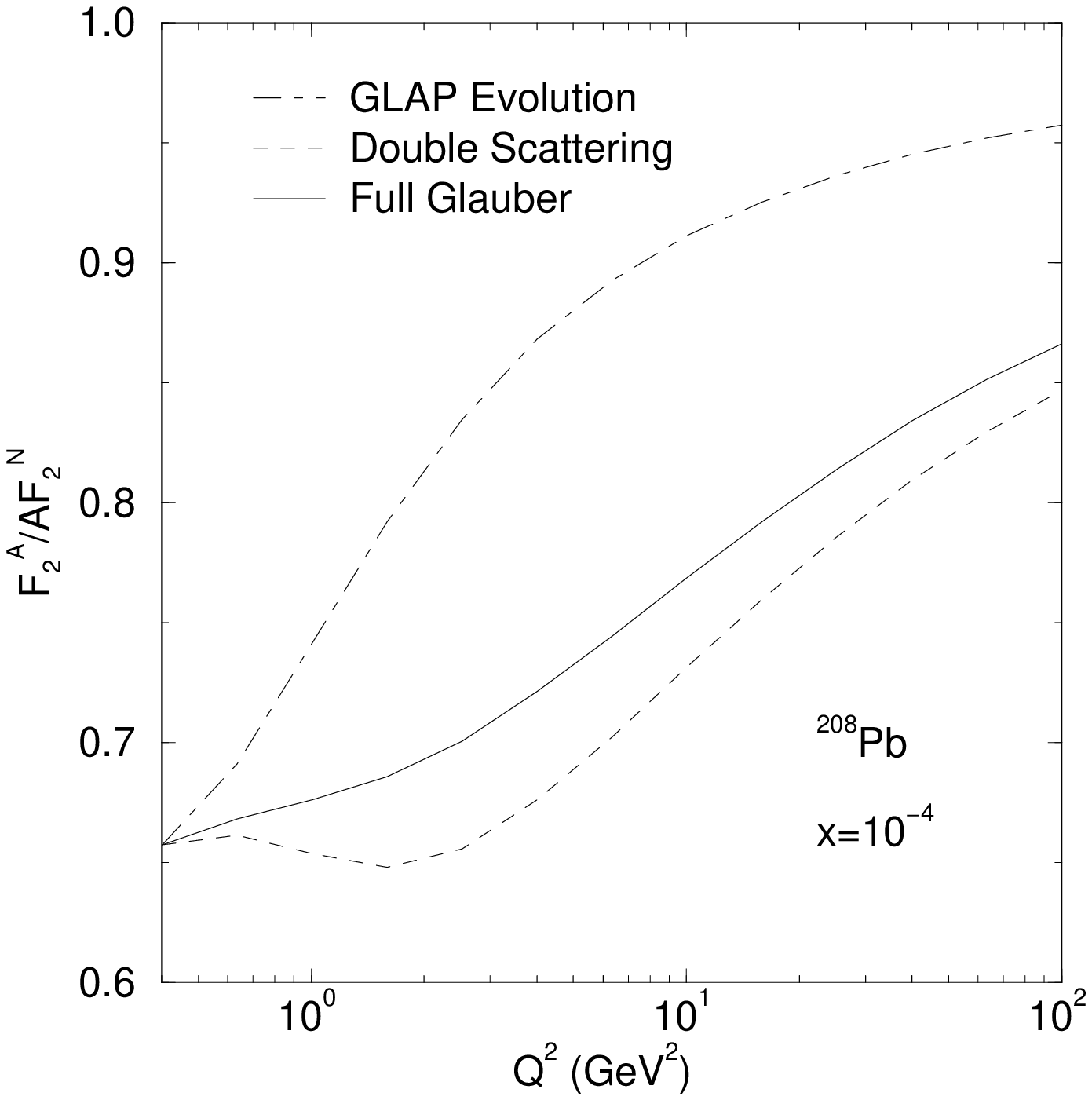}\\
    \rm{Fig.~2}
\nonumber
\end{array}
$$
\end{figure}
We note that at large $Q^2$ the nuclear 
shadowing effect is reduced but does not diminish, i.e. 
it is not a higher twist effect. This can be understood as the interplay
between the perturbative and the non-perturbative shadowing mechanisms, which 
is also evident in the $x$-dependence of $R_q$ \cite{us}.  
The apparent flatness of the 
shadowing ratio at low $Q^2$ in the small-$x$ region is 
altered by the perturbative
evolution. This is due to the singular behavior of $xg_N^{\rm DLA}$
as $x\rightarrow 0$ at large $Q^2$ leading to the strong 
$x$-dependence of the
effective number of scatterings, $\kappa_q(x)$.  Furthermore, at small x 
the 
$Q^2$-evolution of the $F_2^A/A F_2^N$ is much slower than the DGLAP 
evolution due to multiple scatterings.  We illustrate this effect in Fig. 2.  
When 
the effective number of scatterings, $\kappa_q(x)$, is small, the Eq. (3) 
becomes the DLA of the DGLAP evolution equation for quarks.  

Similarly, gluon distribution in a nucleus  
is given by 
\begin{eqnarray}
xg_A(x,Q^2) & = & xg_A(x,Q_0^2)+\frac{2R_A^2}{\pi^2}\int_x^1\frac{dx'}{x'}
\int_{Q_0^2}^{Q^2}
dQ'^2 
 \left [ \gamma +\ln (\kappa_g) +E_1(\kappa_g )
\right ] 
\end{eqnarray}
The essential difference between the quark and gluon cases is 
different splitting functions and the 
effective number of scatterings,  $\kappa_g=9\kappa_q/4$ due to different
color representations that the quark and the gluon belong to. 
These two effects combined result in the 
$12$ times faster increase of the gluon density with $Q^2$ 
than in the case of the sea quarks in the region of small $x$.
The 
two important effects which make the gluon shadowing quite different
from the quark shadowing are the stronger scaling violation in the semihard
scale region and a larger perturbative shadowing effect.  This can be seen 
by considering the shadowing ratio 
\begin{eqnarray}
R_g(x,Q^2) & = & \frac{xg_A(x,Q^2)}{Axg_N(x,Q^2)} 
 =  \frac{xg_N(x,Q_0^2)R_g^0(x)+
\Delta xg_A(x;Q^2,Q_0^2)}{xg_N(x,Q_0^2)+\Delta xg_N(x;Q^2,Q_0^2)}
\end{eqnarray}
where $R_g^0(x)$ is the initial shadowing ratio at $Q_0^2$ and
$\Delta xg(x;Q^2,Q_0^2)$ is the change of the gluon distribution
as the scale changes from $Q_0^2$ to $Q^2$. The strong scaling 
violation due to a larger $\kappa_g$ at small $x$ causes 
$\Delta xg_N(x;Q^2,Q_0^2)\gg xg_N(x,Q_0^2)$ as $Q^2$ is greater than 
$1\sim 2$ GeV$^2$. As seen in Fig. 3(a) for $Q^2\geq 3$GeV$^2$ 
the dependence of $R_g(x,Q^2)$ 
on the initial
condition $R_g^0(x)$ diminishes and the perturbative shadowing
mechanism takes over.  

\begin{figure}[bt]
\vspace{-0.5cm}
$$
\begin{array}{cc}
    \epsfxsize=6.0cm 
    \epsfbox{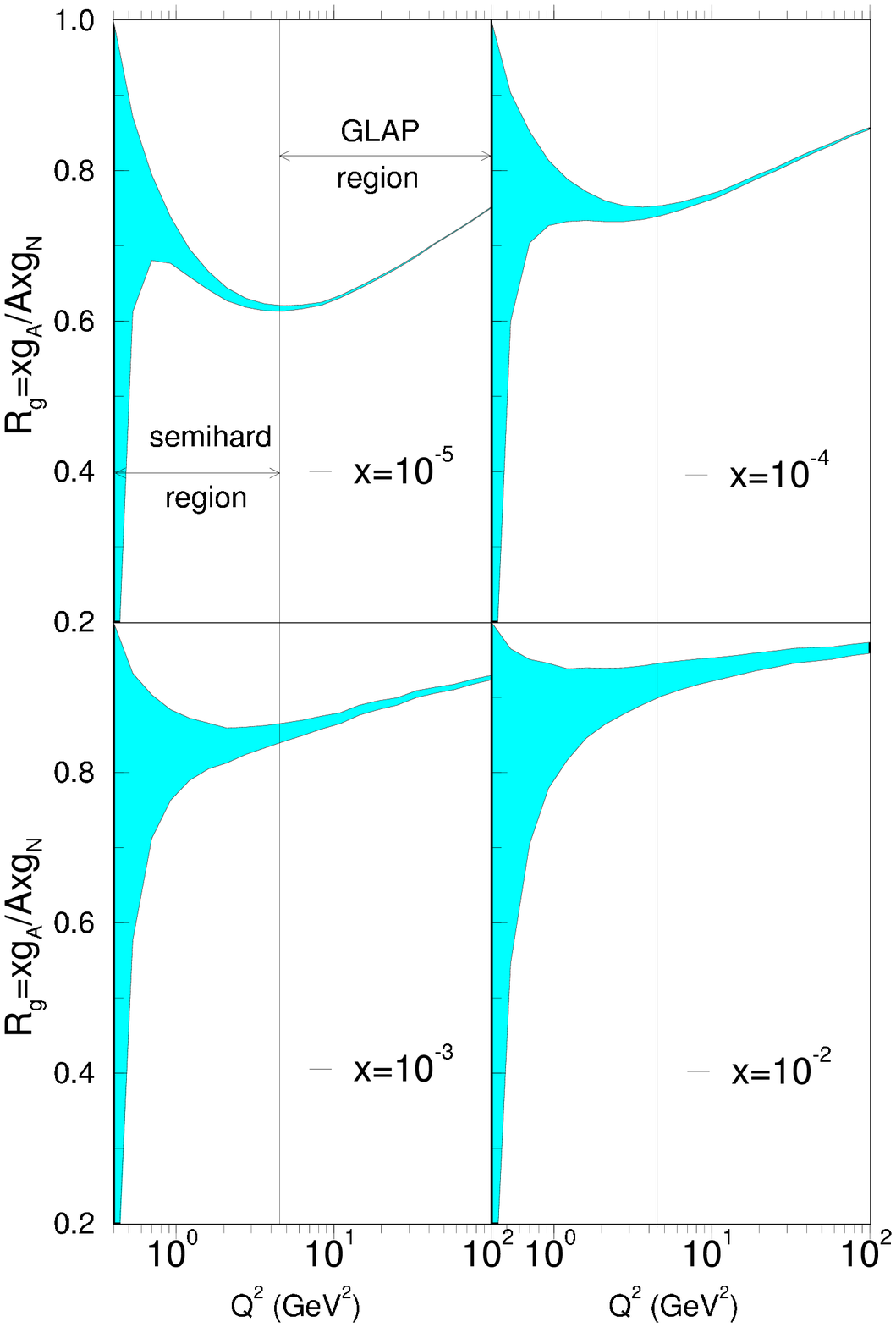} & \!\!\!\!\!\!\!\!\!\!\!\!\!\!\!\!\!\!\! \epsfxsize=9cm \epsfbox{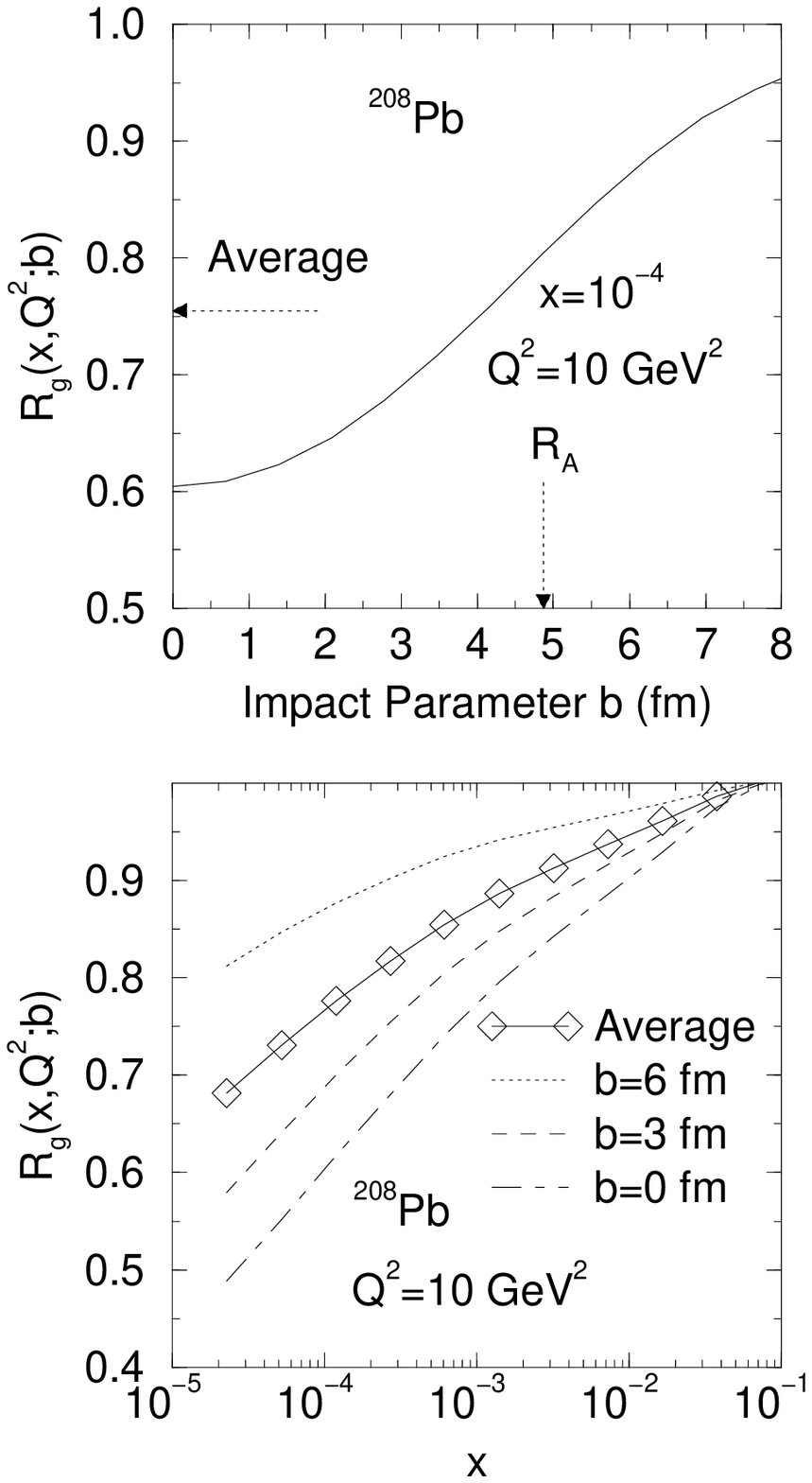}\\
     \\
  \rm{Fig.~3(a)}   &  \rm{Fig.~3(b)} 
\end{array}
\nonumber
$$
\end{figure}

The $x$-dependence of the gluon shadowing can also be predicted as long
as $Q^2>3$GeV$^2$  where the influence of the initial condition is 
minimal. 
The shape of the distribution
 is quite robust in the 
small-$x$ region regardless
of what initial conditions one may choose. Due to the perturbative
nature of the shadowing, these distributions do not exhibit a saturation
as $x$ decreases. 
Furthermore since the shadowing is a non-linear
effect in the effective number of scatterings, 
the impact parameter dependent
shadowing ratio cannot be factorized into a product of
 an average shadowing ratio and the nuclear thickness function.  Our results 
are presented in Fig. 3(b).  

%

In summary, the 
nuclear shadowing phenomenon can be understood as a consequence of the
parton coherent multiple scatterings.
We find that the quark shadowing arises from an interplay between ``soft''
physics and the semihard QCD process.  At small x, we find that the 
quark shadowing ratio evolves with $Q^2$ slower than the DGLAp evolution.  
We show that gluon shadowing is 
largely driven by a perturbative shadowing mechanism due to the strong
scaling violation in the small-$x$ region. 
The gluon shadowing is thus a robust phenomenon at large $Q^2$ and can
be unambiguously predicted by perturbative QCD.
The strong scaling violation of the nucleon structure function in the
semihard momentum transfer region 
at small $x$ can be reliably described
by perturbative QCD and 
is  a central key to the
understanding of the scale dependence of the nuclear shadowing effect.
The impact parameter dependence of gluon shadowing 
is a non-linear effect in the nuclear thickness function. It is
important to correctly incorporate the impact parameter dependence
of the nuclear structure function when one calculates the QCD processes
in the central nuclear collisions.  In the asymptotic limit of small x, 
gluon distribution in nuclei does not exhibit saturation, as in the case 
of the GLR\cite{glr} equation, but rather tends to the limit 
$xG_A \sim R_A^2 Q^2 ln (1/x)$.  Thus, the shadowing ratio, $R_g$, does not 
saturate at small x, as long as the gluon density in a 
nucleon has singular x-dependence at small x.  
\section*{Acknowledgements}  
The work presented here was done in collaboration with Z. Huang and H.J. Lu.  
This work was supported
in part by the U.S. Department of Energy grants 
DE-FG03-93ER40792 and DE-FG03-85ER40213.
%

\end{document}